\documentclass[journal, 10pt, letterpaper]{IEEEtran}
\usepackage[T1]{fontenc}
\usepackage{url}
\ifCLASSINFOpdf
\else
\fi

\usepackage{amsmath}
\usepackage{algorithmicx}
\usepackage{algorithm}
\usepackage{graphicx}
\usepackage{mathtools}
\usepackage{amsthm}
\usepackage{balance}
\usepackage{amssymb}
\usepackage{algpseudocode}
\usepackage{nccmath,textcomp}
\usepackage{eso-pic}
\usepackage{bbm}

\newtheorem*{proof-mine}{Proof}
\makeatletter
\let\MYcaption\@makecaption
\makeatother

\usepackage[font=footnotesize]{subcaption}

\makeatletter
\let\@makecaption\MYcaption
\makeatother
\newcommand\blfootnote[1]{%
  \begingroup
  \renewcommand\thefootnote{}\footnote{#1}%
  \addtocounter{footnote}{-1}%
  \endgroup
}

\begin{document}
\title{Analysis of Non-Pilot Interference on Link Adaptation and Latency in Cellular Networks}
\author{\IEEEauthorblockN{Raghunandan M. Rao, Vuk Marojevic,
Jeffrey H. Reed 
}\\
}
\maketitle
\begin{abstract}
Modern wireless systems such as the Long-Term Evolution (LTE) and 5G New Radio (5G NR) use pilot-aided SINR estimates to adapt the transmission mode and the modulation and coding scheme (MCS) of data transmissions, maximizing the utility of the wireless channel capacity. However, when interference is \emph{localized exclusively on non-pilot resources}, pilot-aided SINR estimates become inaccurate. We show that this leads to congestion due to retransmissions, and in the worst case, outage due to very high block error rate (BLER). We demonstrate this behavior through numerical as well as experimental results with the 4G LTE downlink, which show high BLER and significant throughput detriment in the presence of non-pilot interference (NPI). To provide useful insights on the impact of NPI on low-latency communications, we derive an approximate relation between the \textit{retransmission-induced latency} and BLER. 
Our results show that NPI can severely compromise low-latency applications such as vehicle-to-vehicle (V2V) communications and 5G NR. We identify robust link adaptation schemes as the key to reliable communications. 
\end{abstract}
\blfootnote{
Raghunandan M. Rao and Jeffrey H. Reed are with Wireless@VT, Bradley Department of Electrical and Computer Engineering, Virginia Tech, Blacksburg, VA, 24060, USA (e-mail: \{raghumr,reedjh\}@vt.edu). 

Vuk Marojevic is with the Department of Electrical and Computer Engineering at Mississippi State University, Mississippi State, MS, 39762, USA (e-mail: vuk.marojevic@ece.msstate.edu).

This is the authors' version of the work. For citation purposes, the definitive version of record of this work is: R. M. Rao, V. Marojevic, J. H. Reed, ``Analysis of Non-Pilot Interference on Link Adaptation and Latency in Cellular Networks'', \textit{To Appear in the 89th IEEE Vehicular Technology Conference (IEEE VTC Spring 2019)}, pp. 1-6, April 2019.}

\IEEEpeerreviewmaketitle
\section{Introduction}
LTE has succeeded in meeting its goals and objectives as a global cellular standard since its first release (Release 8). Through its evolution to 5G New Radio (5G NR), it has acquired enhancements across all layers, with new capabilities such as support for vehicular communications \cite{Uhlemann_CV2X_VTM_2017}, device to device (D2D) and the Internet of Things (Narrowband-IoT) \cite{Maldonado_2017_NBIoT}. Because of its success in the commercial and civilian sectors, LTE is being targeted to be used in public safety \cite{Kumbhar_PubSafe_2016}, military networks, and control of the smart grid. 

However, LTE is not without its shortcomings. It has been shown to be vulnerable to protocol-aware attacks across all layers \cite{Lichtman_LTEJS_ComMag_2016},  especially the physical layer \cite{Raghu_MILCOM_2017}, \cite{Vuk_VTC_2017}. A particular class of control channel vulnerabilities that is well known is \emph{pilot/reference signal interference} \cite{Clancy_PilJam_2011}. Since pilot signals are used for channel estimation and coherent demodulation, localized interference on pilot resource elements in OFDM-based systems result in a significantly lower SINR than that in the case of \textit{non-pilot interference}. 

Shahriar et al. \cite{Shahriar_PilRand_2013} proposed and evaluated pilot-tone randomization as a strategy to mitigate pilot interference. The key motivation behind their scheme is to facilitate accurate channel estimation by changing pilot locations in a pseudo-random manner, so that uncorrupted data symbols are decoded and the resulting bit error rate (BER) is much lower than that in pilot interference. Karlsson et al. \cite{Karlsson_TDD_Massive_MIMO_Jam_2017} theoretically analyzed the vulnerability of TDD-massive MIMO to pilot interference. Xu et al. \cite{Xu_Ind_Code_Check_2019} develop a novel coding scheme to authenticate the channel training phase using an independence-checking coding (ICC)-based protocol. In our prior work in \cite{Raghu_MILCOM_2017}, we have demonstrated the vulnerability of LTE to pilot interference through software-defined radio (SDR)-based experiments. 

In this paper, we demonstrate the vulnerability of cellular link adaptation mechanisms to \textit{non-pilot interference} (NPI). 
Link adaptation operates by adapting the transmission mode (e.g. SISO, MIMO, diversity, beamforming) and modulation and coding scheme (MCS) as a function of the SINR. Since modern cellular standards such as LTE and NR implement \textit{pilot-aided SINR estimation}, these estimates will be accurate as long as the interference and noise statistics on the pilots and data resources are the same. Unfortunately, NPI does not obey this condition, since pilots will be interference-free while non-pilots will be interference-corrupted. We analyze the impact of NPI on link adaptation considering the LTE downlink as an example, and observe significant degradation of block error rate (BLER) and throughput through experimental and simulation results. In addition, we define \textit{`retransmission-induced latency'}, a metric that quantifies the latency only due to retransmissions, and derive an approximate expression relating it with the BLER. We develop useful insights through simulation results, comment on the accuracy of our expressions, and highlight the detrimental impact of NPI on low-latency communications. 

The rest of the paper is organized as follows: Section \ref{Sec2_Overview} provides a brief overview of link adaptation and CSI feedback in cellular systems. Section \ref{Sec3_Non_Pil_Interf} introduces the general structure of NPI and its different types. Section \ref{Sec4_Results_Thpt_BLER} describes our experimental setup, simulation methodology and the key results. Section \ref{Sec5_Retx_ind_latency} derives an approximate relation between the BLER and the average retransmission-induced latency $(\bar{\tau}_{\mathtt{retx}})$, and provides the analysis and simulation results of the impact of NPI on $\bar{\tau}_{\mathtt{retx}}$. Section \ref{Sec6_Conc} concludes the paper.

\section{Overview of Link Adaptation and CSI Feedback in Cellular Systems} \label{Sec2_Overview}
Link adaptation is the process of adapting the transmission parameters and hence, the data rate, as a function of the SINR at the receiver. In modern cellular systems, such as 4G LTE and 5G NR, the transmission parameters include the transmission mode (TM) and modulation and coding scheme (MCS). Example of TMs include single-input-single-output (SISO), multiple-input-multiple-output (MIMO), diversity and beamforming. Examples of  MCS include a combination of digital modulation schemes such as BPSK/QPSK/QAM, and different forward error correction (FEC) coding schemes and rates \cite{sesia2011lte}. The channel state information at the receiver (CSIR) is periodically fed back to the transmitter to successfully implement link adaptation schemes. In LTE and NR, the transmitter sends pilot/reference signals to the receiver to help in estimating the CSIR, which is possible since they are known to both the transmitter and reciever. The receiver quantizes the pilot-aided SINR estimate to limit CSI feedback overhead, and this quantization function depends on the cellular standard. Up to Third Generation Partnership Project (3GPP) Release 14, CSI consists of the following parameters: 
\begin{enumerate}
\item  Channel Quality Indicator (CQI): A 4-bit value that is mapped to a 5-bit MCS value.
\item Precoding Matrix Indicator (PMI): An index from a fixed library of precoding matrix elements.
\item Rank Indicator (RI): Index of transmission rank that the user can support.
\end{enumerate}
The 3GPP standard specifies that for the downlink, up to two LTE codewords can be transmitted for each user, where
the MCS value for each codeword must be chosen to keep the $\text{BLER below } 10\%$ \cite{sesia2011lte}. In practical deployments, a lookup table-based approach is used for each TM to map the SINR to a CQI value \cite{Rupp_Sys_LEv_Sim_VTC_2010}.

\begin{figure}[t]
\centering
\includegraphics[width=2.8in]{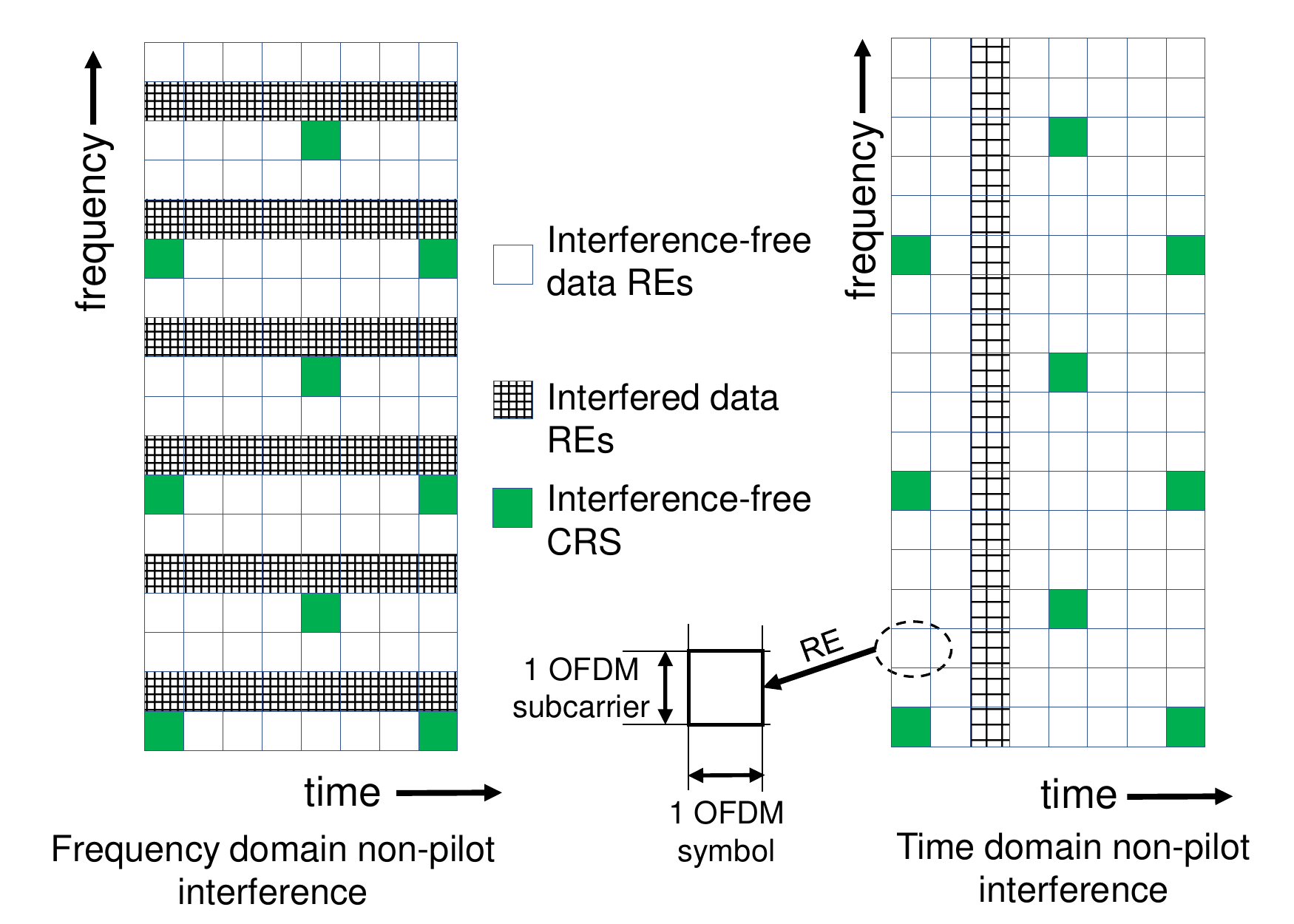}
\caption{Illustration of non-pilot interference (NPI) on the LTE resource grid, and its different types: Frequency-domain NPI and Time-domain NPI.}
\label{Fig1}
\end{figure}

\section{Non-Pilot Interference}\label{Sec3_Non_Pil_Interf}
For pilot-aided CSI estimates, the CSIR is accurate as long as the noise and interference statistics on the pilot and the non-pilot resources are the same. While this is true in typical multi-cellular co-channel interference, it will not be the case in the presence on interference that \textit{only affects non-pilot resources}. We term such interference as \textit{`non-pilot interference'}.  

In typical cellular signals, pilots are sparsely allocated resources in time, frequency and spatial layers. Fig. \ref{Fig1} shows the example of cell-specific reference signals (CRS), which are pilot resource elements in LTE \cite{sesia2011lte}. In later releases of LTE and 5G NR, there are additional pilots such as demodulation reference signals (DMRS) and CSI-Reference Signals (CSI-RS), which are used for coherent demodulation and CSI estimation for more advanced multi-antenna techniques \cite{sesia2011lte}. Similar to CRS, these reference signals are also sparsely located in the OFDM resource grid.

Hence an interferer can leverage the sparsity of pilot resources to localize interference on non-pilot resources, as shown in Fig. \ref{Fig1}, to \textit{contaminate the CSI estimated by the receiver}. In contrast to pilot interference (PI) which contaminates the channel estimates at the receiver \cite{Lichtman_LTEJS_ComMag_2016}, non-pilot interference (NPI) contaminates the CSIR (SINR estimates). Hence, the CSI reports (CQI, PMI and PI) fed back to the transmitter will be inaccurate in the case of NPI.

Based on the localization of power in time and frequency, non-pilot interference can be broadly classified into:
\begin{enumerate}
\item Time-domain non-pilot interference (TD-NPI): The interference can take a pulsed form targeting OFDM symbols between two pilots, as shown in Fig. \ref{Fig1}. Example: On-off jammers, and pulsed-radar signals.
\item Frequency-domain non-pilot interference (FD-NPI): Interference can be localized in between pilot subcarriers, as shown in Fig. \ref{Fig1}. Example: Multi-tone NPI.  
\end{enumerate} 
Unintentional interference such as a pulsed radar signal may not exclusively affect non-pilot resources \textit{all the time}. However, for large pilot spacing (in time) or large radar pulse repetition intervals, the probability of radar affecting non-pilot resources is high. 

\section{Impact of Non-Pilot Interference on LTE}\label{Sec4_Imp_Non_Pil_Interf} \label{Sec4_Results_Thpt_BLER}
\begin{figure}[t]
	\centering
	\includegraphics[width=3.0in]{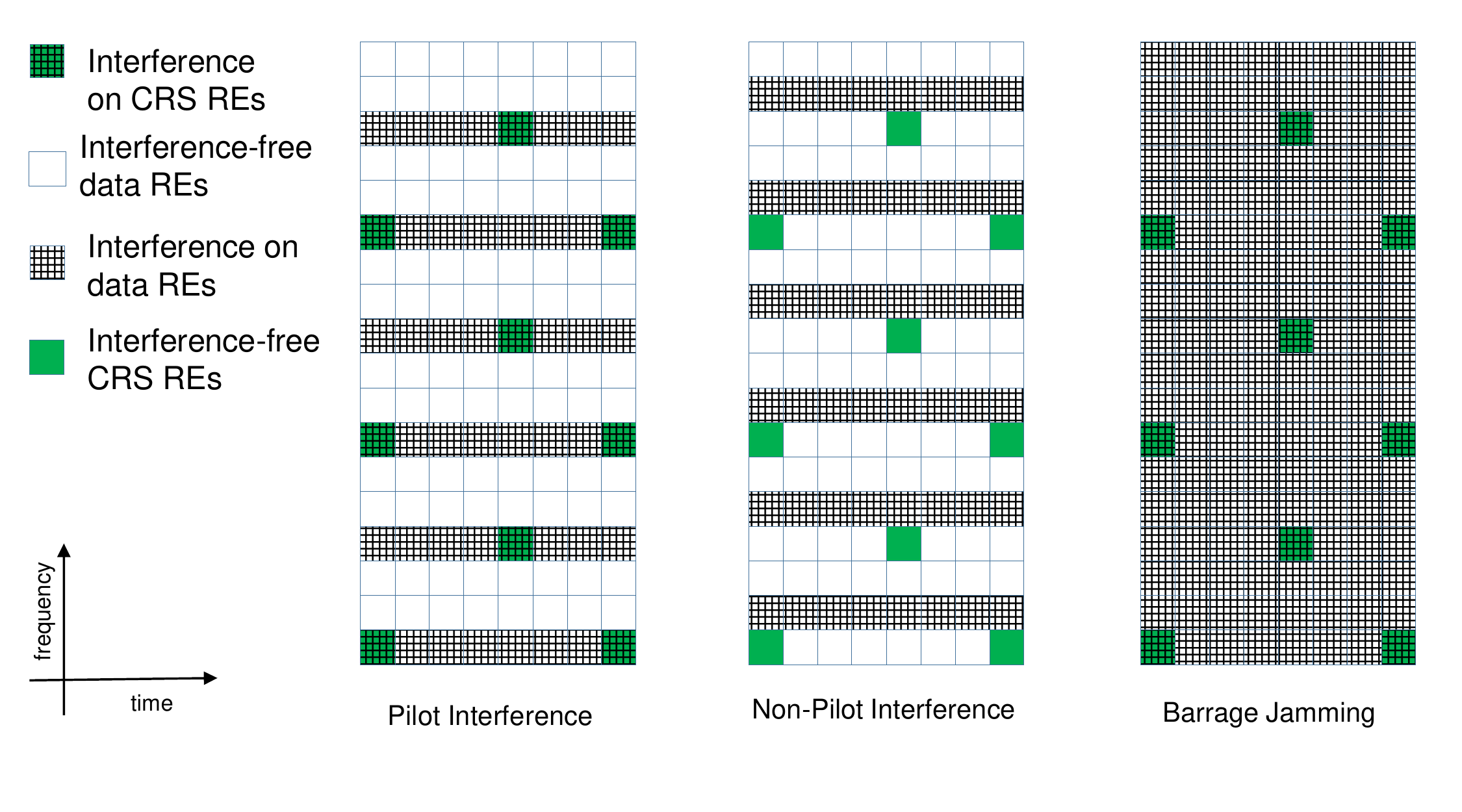}
	\caption{Schematic of the multi-tone interference strategies considered: PI, NPI and barrage jamming. For the same power per subcarrier, barrage jamming needs three times the power than PI and NPI.}
	\label{Fig3_JammingStrategies}
\end{figure}

We consider the LTE downlink with an eNodeB serving a single user equipment (UE) in the presence of multi-tone interference. We consider three scenarios: 
\begin{enumerate}
\item Pilot interference (PI): The interference is localized on pilot subcarriers, and non-pilot subcarriers are unaffected. Such interference has been studied in \cite{Lichtman_LTEJS_ComMag_2016}-\cite{Xu_Ind_Code_Check_2019}.
\item Non-Pilot Interference (NPI): In contrast to pilot interference, the interference is localized exclusively on non-pilot subcarriers. 
\item Barrage Jamming: This is a common technique of wideband interference, where all subcarriers are affected.
\end{enumerate}
These interference strategies are shown in Fig. \ref{Fig3_JammingStrategies}. Prior work on OFDM pilot jamming has shown that the detriment in performance can be mitigated by pseudo-randomly changing the pilot locations to evade the jammer \cite{Clancy_PilJam_2011, Shahriar_PilRand_2013}. It is interesting to note that the resulting scenario is equivalent to NPI. 
As our results will show, evading pilot interference will not provide performance improvements predicted by the theoretical analysis in \cite{Clancy_PilJam_2011}, \cite{Shahriar_PilRand_2013}.

We consider a multi-tone non-pilot interferer with equal spacing between adjacent frequency tones, with equal power allocation across all the targeted OFDM subcarriers. Below, we describe the experimental and simulation methodology to evaluate the impact of these interference strategies.

\begin{figure}[t]
\centering
\includegraphics[width=3.0in]{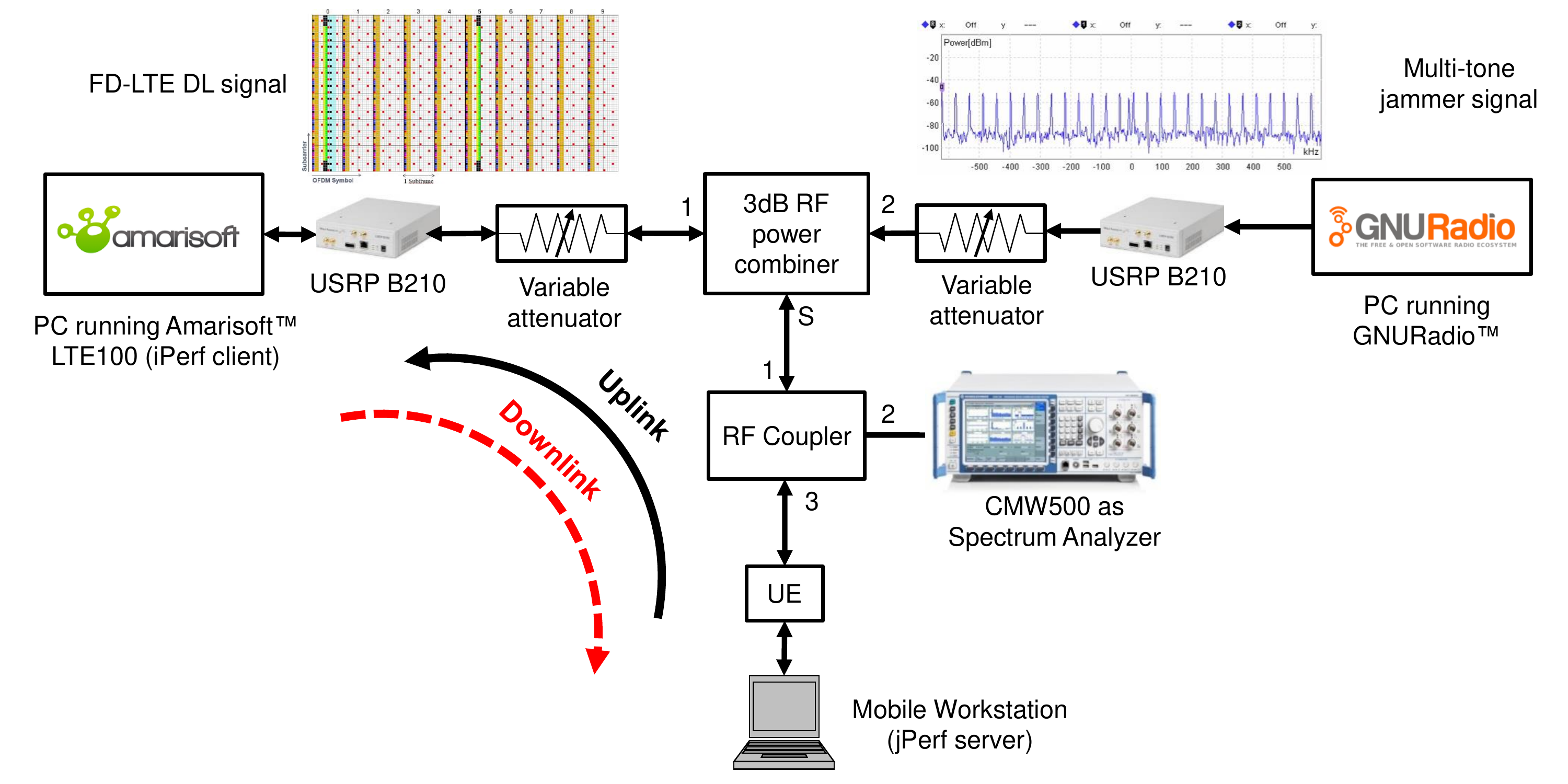}
\caption{Schematic of the LTE downlink multi-tone interference experiments using the Virginia Tech LTE-CORNET Testbed.}
\label{Fig4_Expt_setup}
\end{figure}

\begin{table}[t]
\renewcommand{\arraystretch}{1.0}
\caption{Experimental parameters}
\label{Tab1_experiment_param}
\centering
\begin{tabular}{|l|l|}
\hline
\textbf{Parameter} & \textbf{Description}\\
\hline
LTE Release & 3GPP Release 10\\
\hline 
Frequency band & Band 7 \\
\hline
Bandwidth & $10 \text{ MHz}$ \\
\hline
Reference signal received & -72 dBm \\
power (RSRP) & \\
\hline 
Transmission Mode & TM 0 (SISO) from Port 0 \cite{sesia2011lte} \\
\hline
Channel & Cabled setup with SNR $>30$ dB.\\
\hline 
CSI feedback mode & Periodic and Wideband \\
\hline
CSI periodicity & $10 \text{ ms}$\\
\hline
HARQ mode & Asynchronous with upto 4 retransmissions \\
\hline
\end{tabular} 
\end{table}

\subsection{Experimental Setup}\label{Sec4_Expt_Setup}
The experimental setup to evaluate LTE's performance in multi-tone interference is shown in Fig. \ref{Fig4_Expt_setup}. We used Amarisoft\texttrademark , a proprietary SDR-based LTE eNodeB, and LTE test UEs of Virginia Tech's LTE-CORNET Testbed \cite{Vuk_Rao_WCNC_2017}. We used a cabled setup enclosed in a Faraday cage to isolate the experiment to/from external RF signals. The Amarisoft eNB was interfaced with a Universal Software Radio peripheral (USRP) to generate the LTE downlink signal. Commercial LTE UE dongles are connected to a laptop through a USB interface, to emulate a practical scenario. We monitor the LTE downlink and the jammer spectrum using the CMW 500 Rohde and Schwarz\texttrademark\ LTE test equipment, and control the LTE and multi-tone interference power using variable attenuators as shown in Fig. \ref{Fig4_Expt_setup}. For each SINR value, full buffer traffic was generated using an iPerf client at the PC running Amarisoft, and a jPerf server running on the laptop connected to the LTE UE. We generated custom multi-tone interference waveforms using GNURadio\texttrademark . The experimental parameters are shown in Table \ref{Tab1_experiment_param}. For each SINR value and interference scenario (PI/NPI/barrage), the average throughput and the BLER is measured for $10^5$ frames of LTE downlink traffic to the UE. 

\subsection{Link-level Simulation Study}\label{Link_lvl_sim_study}
We simulated the LTE downlink using a link-level simulator using the MATLAB\texttrademark\ LTE Toolbox, consisting of an eNB serving a single UE in the presence of multi-tone interference in a doubly selective (time and frequency) fading channel. We used the extended pedestrian-A (EPA) channel model to simulate the multipath fading characteristics with a Doppler frequency of $20$ Hz and $\text{SNR} \geq 25$ dB. The rest of the parameters are the same as shown in Table \ref{Tab1_experiment_param}. 
 
\begin{figure}[t]
\centering
\begin{subfigure}[t]{0.5\textwidth}
	\label{3a}
	\centering
	\includegraphics[width=2.8in]{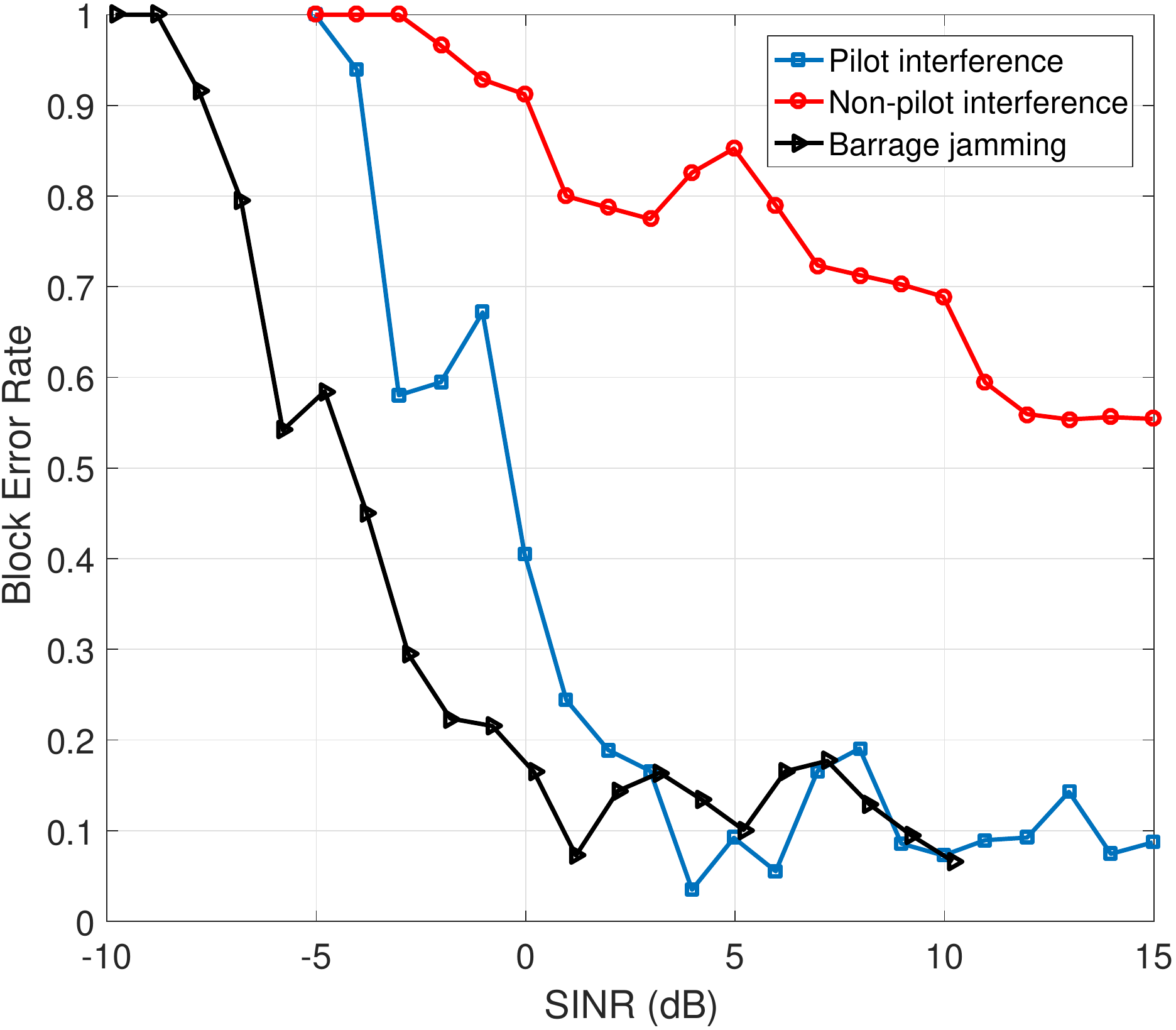}
	\caption{}
\end{subfigure}

 \begin{subfigure}[t]{0.5\textwidth}
	\label{3b}
	\centering
	\includegraphics[width=2.8in]{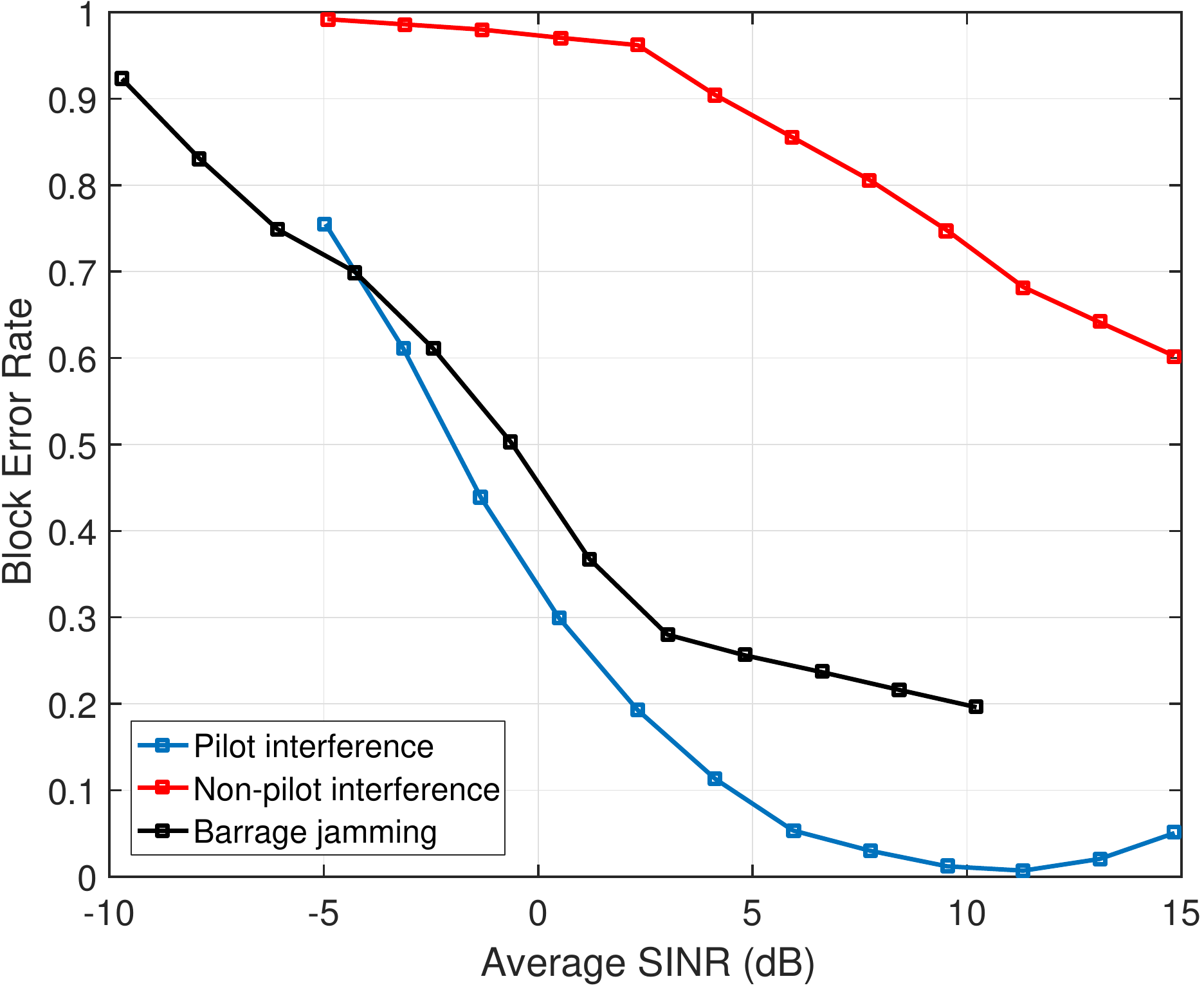}
	\caption{}
\end{subfigure}
\caption{BLER as a function of SINR for PI, NPI and barrage interference, measured during (a) hardware experiments, and (b) link-level simulations.}
\label{Fig6_Expt_BLER_all}
\end{figure}


\begin{figure}[t]
\centering
\includegraphics[width=2.8in]{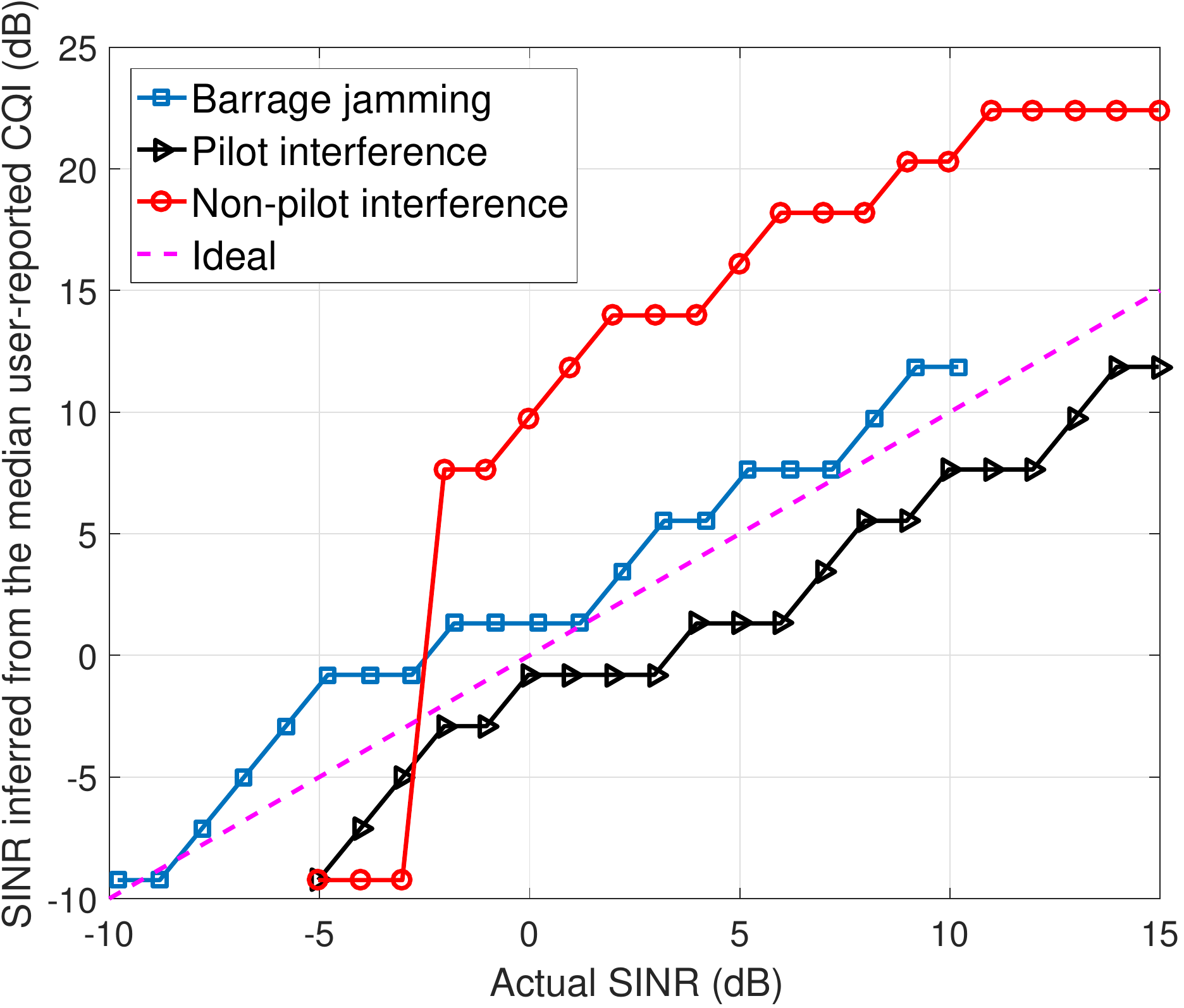}
\caption{Comparison of the actual SINR ($\mathtt{SINR_{act}}$) versus the SINR inferred from the median user-reported CQI ($\mathtt{SINR^{(med)}_{CQI}}$), during our hardware experiments for PI, NPI and barrage interference.}
\label{Fig9_SINR_actual_vs_reported}
\end{figure}

\begin{figure}[t]
\centering
\begin{subfigure}[t]{0.5\textwidth}
	\label{3a}
	\centering
	\includegraphics[width=2.8in]{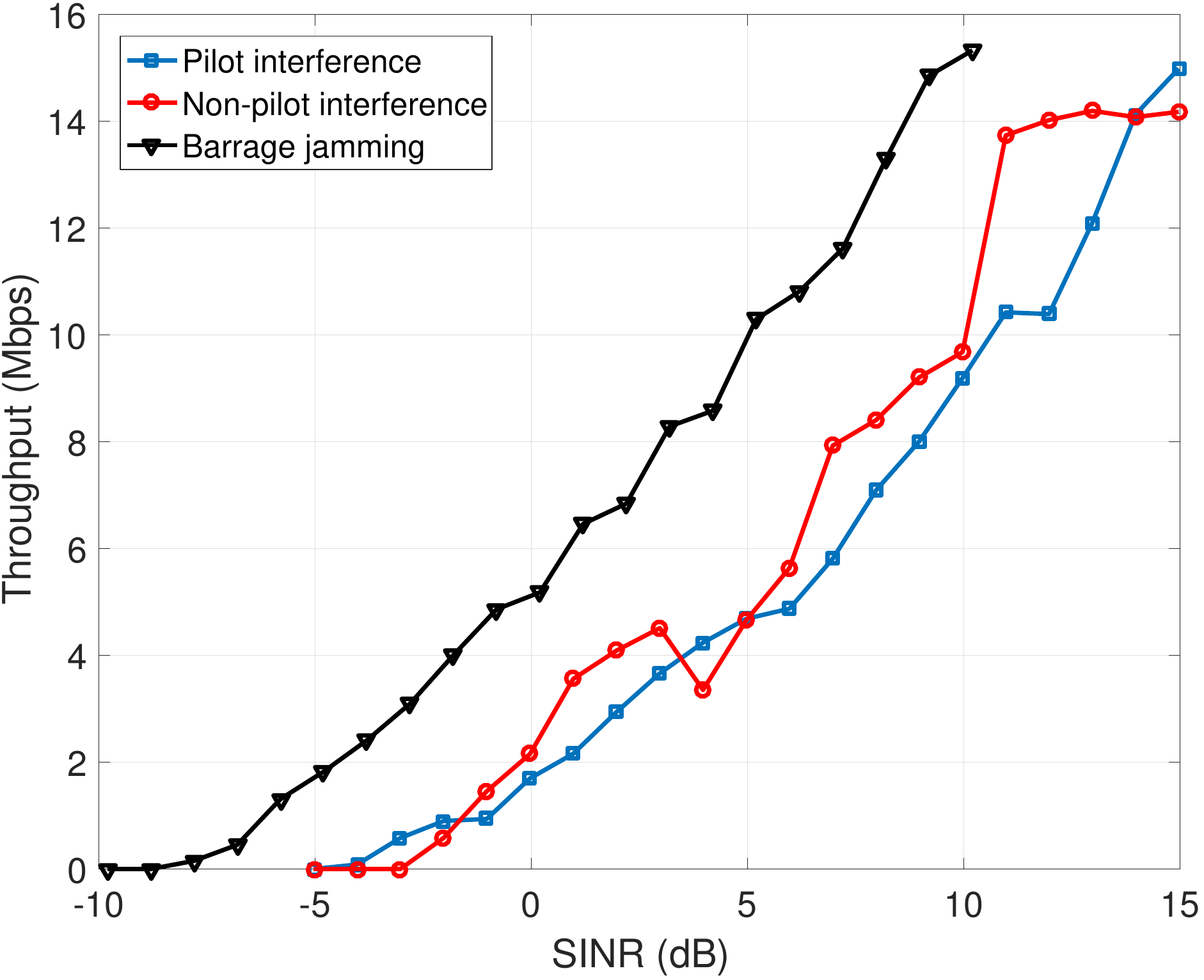}
	\caption{}
\end{subfigure}

\begin{subfigure}[t]{0.5\textwidth}
	\label{3b}
	\centering
	\includegraphics[width=2.8in]{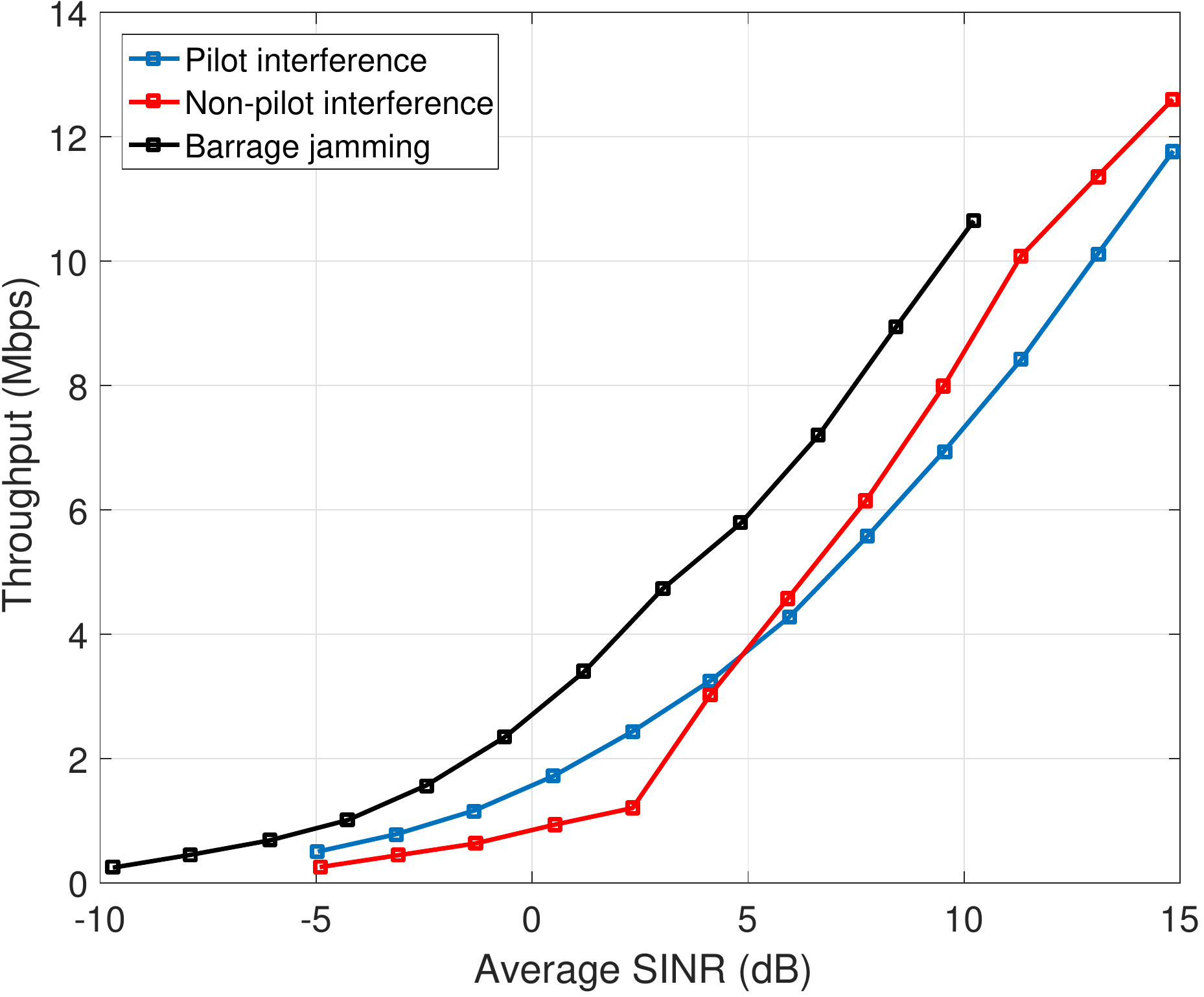}
	\caption{}
\end{subfigure}
\caption{Throughput as a function of SINR for PI, NPI and barrage interference, measured during (a) hardware experiments, and (b) link-level simulations.}
\label{Fig5_Expt_thpt_all}
\end{figure}


\subsection{Results}
In the absence of multi-tone interference, we measured an average throughput of $36.5$ Mbps, which is very close to the theoretical maximum throughput of $37.5$ Mbps achievable with a  10 MHz LTE system. In addition, the measured BLER was consistently under $10 \%$ in the absence of any interference, indicating successful link adaptation. Fig. \ref{Fig6_Expt_BLER_all}(a) shows the measured BLER performance for different multi-tone interference scenarios, in our hardware experiment from Fig. \ref{Fig4_Expt_setup}. We observe that the BLER in non-pilot interference is above $50 \%$ which \textit{indicates link adaptation failure}, since successful link adaptation has a $\text{BLER} \leq 10\%$. On the other hand, the BLER in pilot and barrage interference fluctuates around $10\%$ up to an SINR of $0$ dB, and then gradually begins to deteriorate. We observe a similar trend in Fig. \ref{Fig6_Expt_BLER_all}(b), which shows the BLER performance in our link-level simulations. 

During our hardware experiments described in \ref{Sec4_Expt_Setup}, the actual SINR ($\mathtt{SINR_{act}}$) was measured for each interference scenario using a spectrum analyzer before each downlink transmission. The \textit{SINR estimated by the UEs} was inferred by monitoring the CQI reports sent by the UE to the eNB for all interference scenarios. Most UE manufacturers use a lookup table-based approach to calculate CQI from the pilot-aided SINR estimate. Assuming a SINR-to-CQI mapping given by\footnote{We obtained this mapping from MATLAB's LTE Toolbox. The SINR-to-CQI mapping is not unique, but varies from vendor to vendor. Similar SINR-to-CQI mappings have been reported in the literature, for e.g. in \cite{Rupp_Sys_LEv_Sim_VTC_2010}. } $\mathtt{SINR_{CQI}} \text{ (dB)}= 2.11 \mathtt{CQI} - 9$, the SINR inferred from the median user-reported CQI ($\mathtt{SINR^{(med)}_{CQI}}$) was calculated. Fig. \ref{Fig9_SINR_actual_vs_reported} shows the actual SINR versus the quantized SINR reported by the UE to the eNB in its periodic CQI reports. While $|\mathtt{SINR_{act}} - \mathtt{SINR^{(med)}_{CQI}}| \approx 3 \text{ dB}$ for PI and barrage jamming, for NPI the value of $|\mathtt{SINR_{act}} - \mathtt{SINR^{(med)}_{CQI}}|$ ranges from $7-10$ dB. The steep decline in $\mathtt{SINR^{(med)}_{CQI}}$ at $\mathtt{SINR_{act}} = -3$ dB for NPI is an artifact of the SINR-to-CQI mapping. When the UE experiences outage after detaching from the eNB, the CQI is taken to be 0 by default, for which the corresponding $\mathtt{SINR^{(med)}_{CQI}}=-9$ dB. 

Such overly optimistic SINR estimates with NPI leads to the use of a higher MCS which is not supported by the downlink data channel. Wrongly decoded blocks are retransmitted and when the BLER is high, retransmissions are more frequent. This means that the link adaptation procedure optimized for 3GPP channel models fail when the channel is impaired by multitone interference, especially when pilot signals are not impaired but the data channel is corrupted by interference.

Fig. \ref{Fig5_Expt_thpt_all}(a) shows the measured throughput of the LTE downlink for different interference scenarios. We observe that the throughput performance is close for PI and NPI. However, we observe that outage occurs in NPI at a higher SINR when compared to PI. This behavior can also be observed in our link-level simulation throughput results in Fig. \ref{Fig5_Expt_thpt_all}(b). We also observe that barrage jamming requires about $5$ dB more power than PI/NPI to cause the same throughput detriment. 


\section{Impact of Non-Pilot Interference on Low-latency Communication Systems} \label{Sec5_Retx_ind_latency}
The end-to-end latency in a wireless link encompasses contributions from various sources such as propagation, queuing, scheduling and signal processing. Here, we are particularly interested in the \textit{retransmission-induced latency} ${\tau}_{\mathtt{retx}}$, which we define as the the latency to a user \textit{only due to retransmissions}. The the initial transmission is excluded from this latency metric, so that $\tau_{\mathtt{retx}}=0$ if the initial transmission succeeds.  

To develop useful insights on the impact of NPI on latency, we use the following system model to derive an approximate relation between the $\mathtt{BLER}$ and `average' $\tau_{\mathtt{retx}}$ ($\bar{\tau}_{\mathtt{retx}}$) 
\begin{enumerate}
\item Each user is allocated resources in entities of \textit{blocks}, in which the bits are interleaved and encoded\footnote{In LTE, this is called a transport block which is sent over a transmission time interval (TTI) of 1 ms \cite{sesia2011lte}.}. 
\item For each user, the channel and interference is assumed to be quasi-static. The outcome of each scheduling interval forms an i.i.d. sequence $\mathcal{B}_N=\{X_1,X_2,\cdots,X_N \}$ of Bernoulli trials $X_i$ ($i=1,2,\cdots,N$), each having a \textit{probability of success} $p$. 
\item `Success' is defined as the successful decoding of the data block. 
\item Random process $\mathcal{B}_N$ is \textit{ergodic}. Hence $p=(1-\mathtt{BLER})$ for $N \rightarrow \infty$, since $(1-\mathtt{BLER})$ represents the fraction of blocks successfully decoded.
\end{enumerate}
In reality the channel is non-stationary and non-ergodic due to time-varying SINR. In addition, LTE allows a maximum of 4 HARQ retransmissions. Accounting for these factors is beyond the scope of this paper, and interested readers are referred to \cite{Zhuang_HARQ_Delay_SPecEff_2017} for more details. Below, we demonstrate with our results that the simplified system model is adequate to develop good insights into the impact of NPI on $\bar{\tau}_{\mathtt{retx}}$.

The number of retransmissions $N_{\mathtt{retx}} \in \{0\} \cup \mathbb{N}$, required to successfully decode the transport block is a \emph{geometric random variable}. Therefore, its mean $\bar{N}_{\mathtt{harq}} = \mathbb{E}[N_{\mathtt{harq}}]$ is given by $\bar{N}_{\mathtt{harq}} = \frac{1 - p}{p} = \frac{\mathtt{BLER}}{1 - \mathtt{BLER}}$.
Let ${\tau}_{\mathtt{wait}}$ be the waiting time between consecutive retransmissions to the same user, and $\bar{\tau}_{\mathtt{wait}}$ its average. Then $\mathbb{E}[\tau_{\mathtt{retx}}]=\bar{\tau}_{\mathtt{retx}}$ will be
\begin{align}
\label{tau_HARQ_appendix}
\bar{\tau}_{\mathtt{retx}} = \mathbb{E}[N_{\mathtt{harq}} \tau_{\mathtt{wait}}] = \bar{N}_{\mathtt{harq}} \bar{\tau}_{\mathtt{wait}} = \tfrac{\mathtt{BLER} \times \bar{\tau}_{\mathtt{wait}}}{1 - \mathtt{BLER}}.
\end{align}
We notice that if $\mathtt{BLER} \rightarrow 1$, then $\bar{\tau}_{\mathtt{retx}} \rightarrow \infty$ always. In contrast, for perfect link adaptation with $\mathtt{BLER}=0$, we have $\bar{\tau}_{\mathtt{retx}} = 0$. 4G LTE and 5G NR specify that the TM and MCS should be chosen such that $\mathtt{BLER} \leq 10\%$. Substituting this target $\mathtt{BLER}$ in (\ref{tau_HARQ_appendix}), we get $\bar{\tau}_{\mathtt{retx}} \leq 0.89 \text{ ms}$.

\begin{figure}[t]
\centering
\includegraphics[width=2.8in]{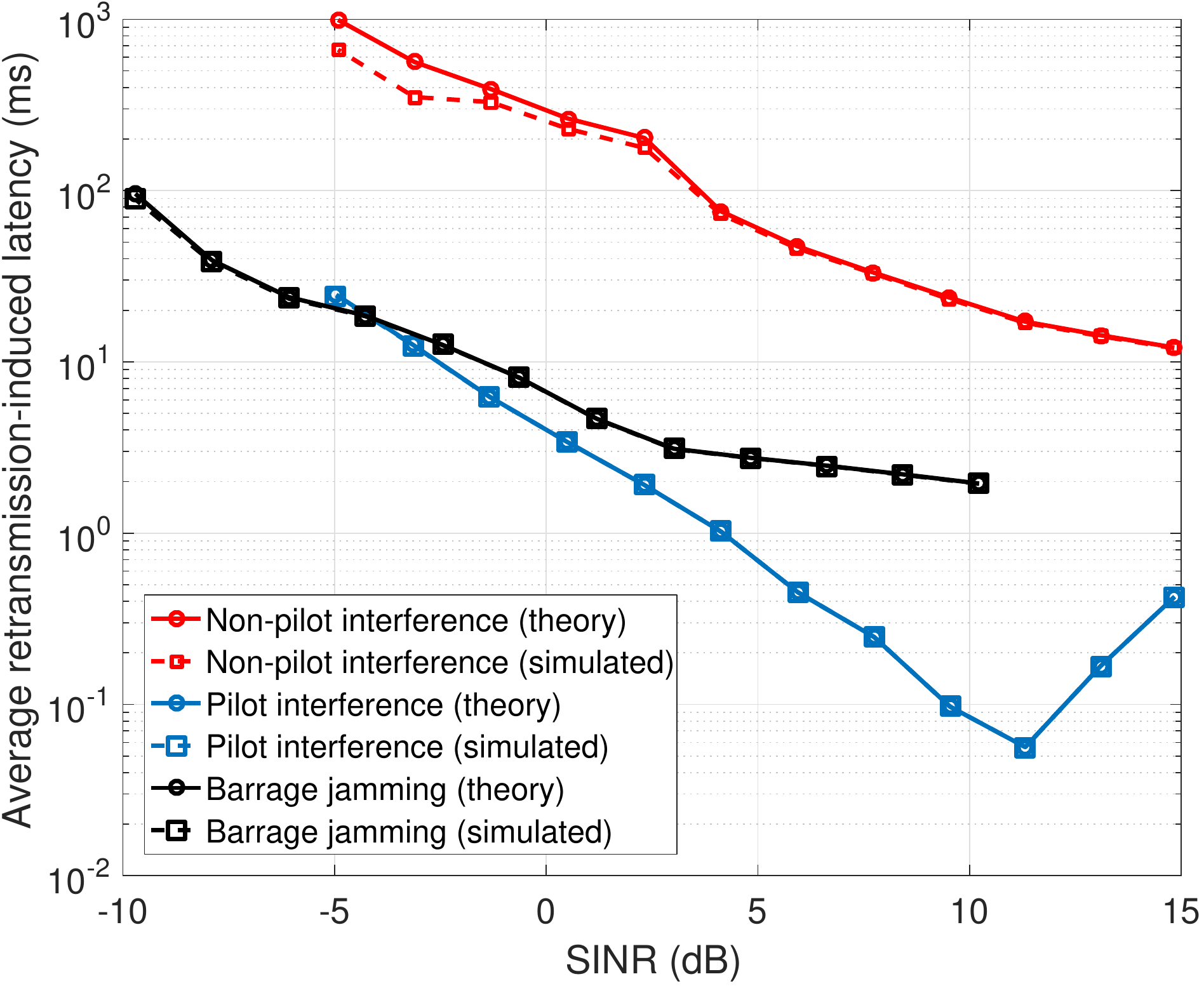}
\caption{Comparison of theoretical (solid lines) versus numerical (dashed lines) values of $\bar{\tau}_{\mathtt{retx}}$ as a function of SINR for PI, NPI and barrage interference. The numerical values are obtained from our link-level simulations.}
\label{Fig10_Retx_induced_latency_all}
\end{figure}

Fig. \ref{Fig10_Retx_induced_latency_all} shows the variation of theoretical and simulated values of $\bar{\tau}_{\mathtt{retx}}$ with SINR, for different interference scenarios. We considered $\bar{\tau}_{\mathtt{wait}}= 8\text{ ms}$ in our link-level simulations, which is the typical round-trip time in LTE \cite{sesia2011lte}. We observe that there is a good agreement between the theoretical and simulated values of $\bar{\tau}_{\mathtt{retx}}$. In the case of PI, the non-monotonic behavior of $\bar{\tau}_{\mathtt{retx}}$ is a direct consequence of the non-monotonic behavior of $\mathtt{BLER}$ in Fig. \ref{Fig6_Expt_BLER_all}(b). Our model overestimates $\bar{\tau}_{\mathtt{retx}}$ for high values of BLER, which is due to the assumption of ergodicity. The intuition behind this trend is the following. Since the instantaneous BLER can vary due to time fading in the channel, a high BLER in the denominator of (\ref{tau_HARQ_appendix}) makes $\bar{\tau}_{\mathtt{retx}}$ very sensitive to perturbations in the SINR. However for low to moderate values of BLER, our results suggest that ergodicity is a reasonable approximation, since small perturbations in $\mathtt{BLER}$ won't cause large deviations in $\bar{\tau}_{\mathtt{retx}}$. 

We observe that NPI has an order of magnitude higher $\bar{\tau}_{\mathtt{retx}}$ when compared to the other interference scenarios. It results in additional latencies of 10-1000 ms for the considered SINRs, whereas a balanced 4G LTE system would have $\bar{\tau}_{\mathtt{retx}} < 1 \text{ ms}$, irrespective of the SNR. Even for high SINR in NPI, $\bar{\tau}_{\mathtt{retx}} \geq 10$ ms which is detrimental for low-latency applications, particularly vehicle-to-vehicle communications and ultra-reliable low latency communications in 5G NR.

\section{Conclusion} \label{Sec6_Conc}
In this paper, we highlighted and analyzed the problem of link adaptation failure in cellular systems due to non-pilot interference (NPI). Using LTE as an example, we demonstrated through experiments and link-level simulations that pilot-aided SINR estimates in NPI are inaccurate and severely degrades link adaptation performance, especially the BLER and throughput. We also derived approximate expressions that relate BLER and \emph{retransmission-induced latency}. Our results indicate that the links affected by NPI becomes unreliable for vehicular communications and low-latency applications of 4G LTE and 5G NR. Further research into robust SINR estimation and link adaptation is necessary to mitigate this problem in current and future cellular networks, and applications such as virtual/augmented reality and connected and autonomous vehicles.
\balance
\bibliographystyle{IEEEtran}
\bibliography{references_wcl}
\ifCLASSOPTIONcaptionsoff
  \newpage
\fi

\end{document}